\documentclass[aps,reprint, prx, superscriptaddress, showpacs]{revtex4-1}
\usepackage{graphicx}
\usepackage{bm,amsmath}
\usepackage{hyperref}
\usepackage{dcolumn}
\usepackage{natbib,xcolor}
\usepackage[version=4]{mhchem}

\begin{document}
\title{Elastocaloric signatures of symmetric and antisymmetric strain-tuning of quadrupolar and magnetic phases in \ce{DyB2C2}}

\author{Linda Ye}
\email{lindaye0@stanford.edu}
\affiliation{Geballe Laboratory for Advanced Materials, Stanford University, Stanford, California 94305, USA}
\affiliation{Department of Applied Physics, Stanford University, Stanford, California 94305, USA}
\author{Yue Sun}
\affiliation{Department of Physics, University of California, Berkeley, California 94720, USA}
\affiliation{Materials Science Division, Lawrence Berkeley National Laboratory, Berkeley, California 94720, USA}
\author{Veronika Sunko}
\affiliation{Department of Physics, University of California, Berkeley, California 94720, USA}
\affiliation{Materials Science Division, Lawrence Berkeley National Laboratory, Berkeley, California 94720, USA}
\author{Joaquin F. Rodriguez-Nieva}
\affiliation{Department of Physics, Stanford University, Stanford, California 94305, USA}
\author{Matthias S. Ikeda}
\affiliation{Geballe Laboratory for Advanced Materials, Stanford University, Stanford, California 94305, USA}
\affiliation{Department of Applied Physics, Stanford University, Stanford, California 94305, USA}
\author{Thanapat Worasaran}
\affiliation{Geballe Laboratory for Advanced Materials, Stanford University, Stanford, California 94305, USA}
\affiliation{Department of Applied Physics, Stanford University, Stanford, California 94305, USA}
\author{Matthew E. Sorensen}
\affiliation{Geballe Laboratory for Advanced Materials, Stanford University, Stanford, California 94305, USA}
\affiliation{Department of Physics, Stanford University, Stanford, California 94305, USA}
\author{Maja D. Bachmann}
\affiliation{Geballe Laboratory for Advanced Materials, Stanford University, Stanford, California 94305, USA}
\affiliation{Department of Applied Physics, Stanford University, Stanford, California 94305, USA}
\author{Joseph Orenstein}
\affiliation{Department of Physics, University of California, Berkeley, California 94720, USA}
\affiliation{Materials Science Division, Lawrence Berkeley National Laboratory, Berkeley, California 94720, USA}
\author{Ian R. Fisher}
\affiliation{Geballe Laboratory for Advanced Materials, Stanford University, Stanford, California 94305, USA}
\affiliation{Department of Applied Physics, Stanford University, Stanford, California 94305, USA}
\date{\today}
\begin{abstract}
The adiabatic elastocaloric effect measures the temperature change of a given system with strain and provides a thermodynamic probe of the entropic landscape in the temperature-strain space. In this study we present an experimental demonstration that the DC bias strain-dependence of AC elastocaloric effect can be employed to decompose the latter into effects from symmetric (rotation-symmetry-preserving) and antisymmetric (rotation-symmetry-breaking) strains, using a tetragonal $f$-electron system \ce{DyB2C2}--whose antiferroquadrupolar order locally breaks four-fold rotational site symmetries while globally remaining tetragonal--as a showcase example. We capture the strain evolution of the quadrupolar and magnetic phase transitions in the system using both singularities in the elastocaloric coefficient $dT/d\epsilon_{xx}$ and its jump at the phase transitions, and the latter we show follows a modified Ehrenfest relation. We find that antisymmetric strain couples to the underlying order parameter in a bi-quadratic manner in the antiferroquadrupolar (AFQ) phase but in a linear-quadratic manner in the canted antiferromagnetic (CAFM) phase, and we attribute the contrast to a preserved (broken) tetragonal symmetry in the quadrupolar (magnetic) phase, respectively. The broken tetragonal symmetry at the CAFM phase is further supported by elastocaloric strain-hysteresis and observation of two sets of domains with mutually perpendicular principal axes in optical birefringence. Additionally, when the quadrupolar moments are ordered in a staggered fashion, we uncover an elastocaloric response that reflects a quadratic increase of entropy with antisymmetric strain, analogous to the role magnetic field plays for Ising antiferromagnetic orders by promoting pseudospin flips. Our results demonstrate AC elastocaloric effect as a compact and incisive thermodynamic probe into the coupling between electronic degrees of freedom and strain, which can potentially be applied to broader classes of quantum materials. 
\end{abstract}
\maketitle

\section{\NoCaseChange{Introduction}}

Strain--and the associated modification of lattice parameters--has long been used as a highly effective means to tune material properties, particularly in the context of hydrostatic pressure \cite{olsen1964superconductivity}. Recent developments of piezoelectric-based devices capable of applying uniaxial stress to materials in a nearly continuous fashion, have highlighted the unique roles that anisotropic strain can play \cite{hicks2014piezoelectric,hicks2014strong,steppke2017strong}. Its application in strongly correlated electron systems where competing phases are expected to be sensitively tuned by external control parameters \cite{dagotto2005complexity} opens up particularly exciting possibilities; in this context, the potential of anisotropic strain as an effective tuning parameter has begun to be demonstrated for a number of superconducting \cite{hicks2014strong,steppke2017strong,malinowski2020suppression,kostylev2020uniaxial}, nematic \cite{kostylev2020uniaxial,worasaran2021nematic} and charge density wave states \cite{kim2018uniaxial,RTe3,YBCO_strain}. Uniaxial stress, like hydrostatic pressure, couples to these phases by modifying the atomic spacing in the lattice hosts; however, an important difference between the two as material tuning parameters is that hydrostatic pressure should, in principle, preserve the space group symmetry, whereas strain induced by uniaxial stress can modify the space group with relatively small lattice distortions.

Whether and how the generated lattice deformation conforms to the symmetries of the pristine crystal structure depends on how strain/stress is implemented in given experiments. In this context, group theoretical irreducible representations provide a basis to decompose an arbitrary infinitesimal deformation into a superposition of orthogonal modes categorized by how they transform under certain symmetry elements in the original, undeformed point group \cite{nye1985physical,luthi2007physical}. This notion of symmetry decomposition has been applied to a number of experimental strain studies in recent years: the critical temperatures of the superconducting transition in \ce{Sr2RuO4} \cite{hicks2014strong} and the nematic transition in iron-based superconductors \cite{ikeda2018symmetric,worasaran2021nematic} have been demonstrated to be tuned by strains belonging to different irreducible representations in highly distinct manners; categorizing strain-induced resistance changes (elastoresistance) also into irreducible representations allows the introduction of elastoresistance tensor \cite{shapiro2015symmetry}, whose components in distinct symmetry channels enabled distinguishing different microscopic mechanisms of coupling between strain and the underlying degrees of freedom \cite{wiecki2020dominant,URu2Si2_wang2020,YbRu2Ge2}. These works demonstrate that decomposing an applied strain into distinct symmetry channels provides an organizing principle towards a systematic understanding of experimentally obtained strain-responses.

In the present work, adding to the symmetry-resolving capabilities of anisotropic strain, we apply a symmetry-decomposition framework to a strain-based thermodynamic quantity, namely, the adiabatic elastocaloric coefficient. We employ this technique to unravel the interplay between symmetric and antisymmetric strains and the underlying anisotropic $f$-electron degrees of freedom in a tetragonal intermetallic \ce{DyB2C2}. The article is organized as follows: we first discuss the elastocaloric effect itself and its symmetry properties in section II; in sections III and IV, we use the elastocaloric effect to categorize the symmetric and antisymmetric strain effects to the quadrupolar and magnetic phase transitions in the system; we then focus on the antisymmetric strain responses away from the phase transitions in section V and in section VI discuss broader scopes of using the symmetry decomposition aspect of the elastocaloric effect as an experimental probe of spatially anisotropic orders and fluctuations. 

\begin{figure}[t]
	\includegraphics[width = \columnwidth]{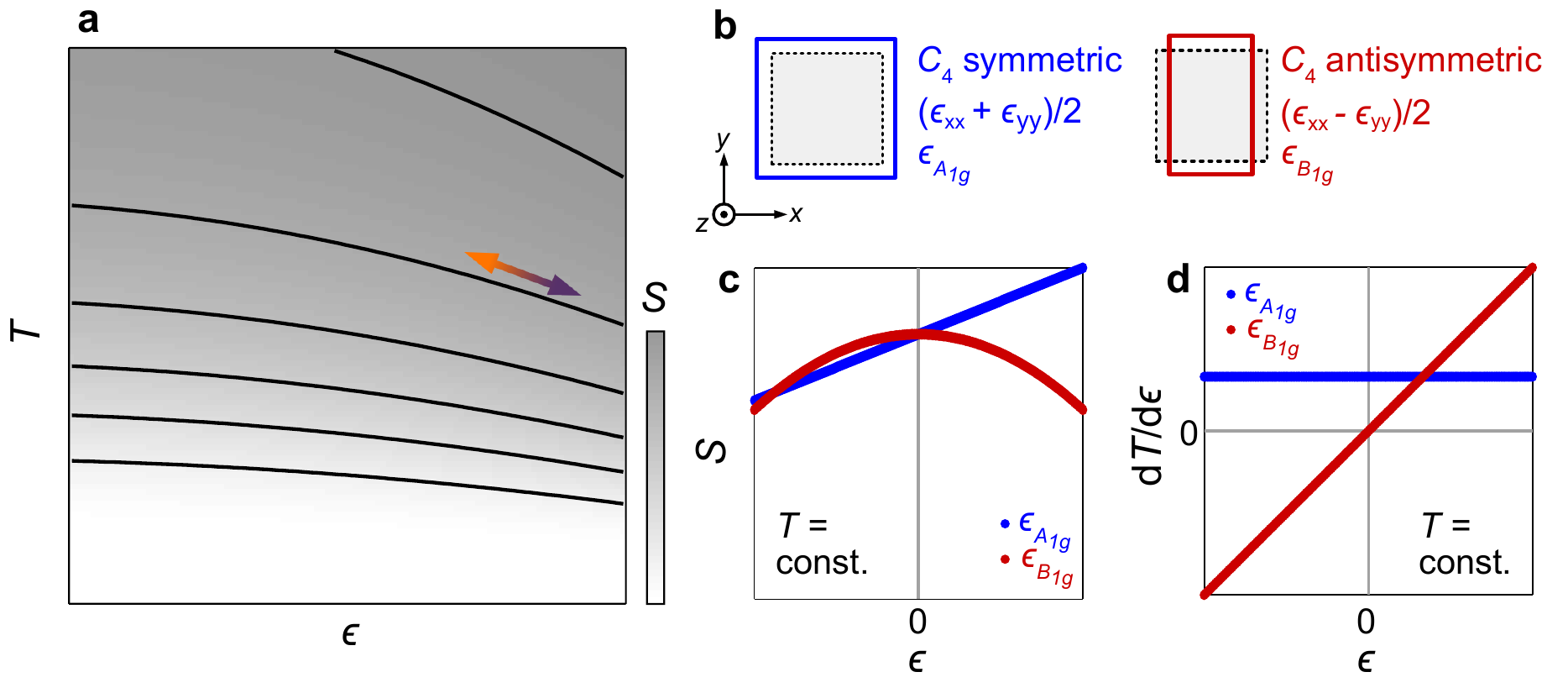}
	\caption{\label{fig-schematic} \textbf{Adiabatic elastocaloric effect and its symmetry properties} (a) Schematic of adiabatic AC elastocaloric effect in a temperature ($T$)-strain ($\epsilon$) phase space, where the black solid lines are isentropic contours. The intensity of gray shade indicates the value of entropy $S$. The colored arrow illustrates the temperature oscillation induced by AC strain along the isentropic lines.  (b) Schematic of $\frac{1}{2}(\epsilon_{xx}+\epsilon_{yy})$ ($\epsilon_{A_{1g}}$, blue box) and $\frac{1}{2}(\epsilon_{xx}-\epsilon_{yy})$ ($\epsilon_{B_{1g}}$, red box). The gray dashed box indicate an undeformed tetragonal lattice. (c,d) Schematic of the lowest-order evolution of $S$ (c) and $dT/d\epsilon$ (d) with $\epsilon$ in the respective symmetric $\epsilon_{A_{1g}}$ (blue) and antisymmetric $\epsilon_{B_{1g}}$ (red) strain channels (see text). }
\end{figure}

\section{\NoCaseChange{Elastocaloric effect and its symmetry decomposition}}

The adiabatic elastocaloric coefficient is defined as the temperature $T$ change of a given system in response to strain $\epsilon$ as $\left({\partial T}/{\partial\epsilon}\right)_S$, and it has recently emerged as an incisive tool for the study of strain-responses of a number of quantum materials \cite{ikeda_ECE,Sr2RuO4_ECE}. In these studies the elastocaloric coefficients are experimentally probed through the measurement of a temperature oscillation in response to an AC strain, at a frequency such that a quasi-adiabatic condition is achieved \cite{ikeda_ECE}. To illustrate the physical origin of the elastocaloric effect, in Fig. \ref{fig-schematic}(a) we depict with a thick two-headed arrow the thermodynamic trajectory of an AC elastocaloric measurement on a generic system in the $T$-$\epsilon$ phase space. As adiabatic processes are confined along the black solid isentropic contours, the normalized temperature oscillation (adiabatic elastocaloric coefficient) can be related to the $\epsilon$-derivative of entropy $S$ in the isothermal condition \cite{ikeda_ECE}:
\begin{equation}\label{dSde}
\left(\dfrac{\partial T}{\partial\epsilon}\right)_S=-\dfrac{T}{C_\epsilon}\left(\dfrac{\partial S}{\partial\epsilon}\right)_T
\end{equation}
Here $C_{\epsilon}$ is the heat capacity at fixed $\epsilon$. Eq. \ref{dSde} implies that the elastocaloric effect can be in turn used to characterize the entropy landscape with $\epsilon$. We note that strictly speaking strain $\epsilon_{ij}$ is a $3\times3$ tensor and in Eq. \ref{dSde} we use $\epsilon$ to denote the experimentally generated linear combinations of $\epsilon_{ij}$ components; the $\epsilon$-derivatives $(\partial T/\partial\epsilon)_S$ and $(\partial S/\partial\epsilon)_T$ are defined for all other components of $\epsilon_{ij}$ being held to zero. We return to the specific combination of strain tensor components shortly. Eq. \ref{dSde} further suggests that $(\partial T/\partial\epsilon)_S\cdot T^{-1}$ may be viewed as a Gr\"{u}neisen parameter with respect to $\epsilon$, which we may denote as $\Gamma_{\epsilon}$, analogous to that defined in the context of hydrostatic pressure and magnetic field \cite{gegenwart2016gruneisen}.

In the following we lay out how spatial symmetry operations constrain the form of the elastocaloric coefficient $\left({\partial T}/{\partial\epsilon}\right)_S$, using the four-fold rotational symmetry $C_4$ as an example. For a system with $C_4$ as one of its symmetry elements (we take $z$ as the four-fold axis and assume that for simplicity the system preserves inversion symmetry and mirror symmetry as well, forming the point group $D_{4h}$), the irreducible representations of strains are $\frac{1}{2}(\epsilon_{xx}+\epsilon_{yy})$, $\epsilon_{zz}$ ($A_{1g}$) and $\frac{1}{2}(\epsilon_{xx}-\epsilon_{yy})$ ($B_{1g}$), $\epsilon_{xy}$ ($B_{2g}$) and $\epsilon_{xz},\epsilon_{yz}$ ($E_g$). In the following we restrict ourselves to normal strains and the shear strain components $\epsilon_{xz},\epsilon_{yz},\epsilon_{xy}$ are kept zero. In Fig. \ref{fig-schematic}(b) we contrast two of the in-plane strain modes $\frac{1}{2}(\epsilon_{xx}+\epsilon_{yy})$ and $\frac{1}{2}(\epsilon_{xx}-\epsilon_{yy})$, which belongs to $A_{1g}$ and $B_{1g}$ representations of $D_{4h}$, respectively; under a $\pi/2$-rotation ($x\rightarrow y,y\rightarrow -x$), the former is symmetric ($\frac{1}{2}(\epsilon_{xx}+\epsilon_{yy})\rightarrow\frac{1}{2}(\epsilon_{xx}+\epsilon_{yy})$) while the latter is antisymmetric ($\frac{1}{2}(\epsilon_{xx}-\epsilon_{yy})\rightarrow-\frac{1}{2}(\epsilon_{xx}-\epsilon_{yy})$). Since $S$ as a scalar remains invariant under $C_4$, the lowest order allowed dependence of $S$ on $\epsilon_{A_{1g}}$($\epsilon_{B_{1g}}$) is linear (quadratic) in a Taylor expansion:
\begin{equation}\label{S_decompose}
S=S^0+S^{A_{1g}^1}\epsilon_{A_{1g}}+S^{A_{1g}^2}\epsilon_{A_{1g}}^2+S^{B_{1g}^2}\epsilon_{B_{1g}}^2+...
\end{equation}
Here $S^{X^i}$ are $\epsilon$-independent coefficients for $\epsilon_X^i (X=A_{1g},B_{1g})$ and in Eq. \ref{S_decompose} we only enumerate up to quadratic terms. These contributions give $\partial T/\partial\epsilon$ the following form:
\begin{equation}\label{ECE_decompose}
\left(\dfrac{\partial T}{\partial\epsilon}\right)_S=D^{A_{1g}^0}+D^{A_{1g}^1}\epsilon_{A_{1g}}+D^{B_{1g}^1}\epsilon_{B_{1g}}+...
\end{equation}
where $D_{X^i}$ are $\epsilon$-independent coefficients. From Eq. \ref{ECE_decompose} we see that an $\epsilon$-independent, constant response in $(\partial T/\partial\epsilon)_S$ necessarily arise from $\epsilon_{A_{1g}}$, while an $\epsilon$-linear contribution may either originate from $\epsilon_{A_{1g}}$ or $\epsilon_{B_{1g}}$. 

In many material systems, Eqs. \ref{S_decompose} and \ref{ECE_decompose} may be further simplified via comparison with hydrostatic pressure experiments which exclusively probe $A_{1g}$ effects. In particular, the strain-dependence of critical temperature $T_c$ of given phase transitions can often be regarded as an indicator of wider free energy landscape; when $T_c$ evolves linearly with pressure, $\epsilon_{A_{1g}}^2$ terms in thermodynamic variables and $\epsilon_{A_{1g}}$-linear term in $(\partial T/\partial \epsilon)_S$ can reasonably be excluded for proximate regions in the $\epsilon-T$ plane \footnote{We note that a linear-dependence of $T_c$ with pressure may also arise from a mutual cancellation between quadratic terms of the two independent $A_{1g}$ components $\frac{1}{2}(\epsilon_{xx}+\epsilon_{yy})$ and $\epsilon_{zz}$, while this requires an unlikely degree of fine-tuning. As a second note, $T_c$ only puts constraints on the free energy close to phase transitions; we hypothesize that when $T_c$ is not strongly modified by $\epsilon$ and $\epsilon_{A_{1g}}^2$ terms are excluded near the phase transition, the presence of such terms in the free energy away from $T_c$ is also unlikely.}.  Under such circumstances including the subject of this study \cite{DyB2C2_pressure}, we may attribute the $\epsilon$-quadratic and -linear components in $S$ to antisymmetric and symmetric channels illustrated respectively in red and blue in Fig. \ref{fig-schematic}(c). The corresponding constant and $\epsilon$-linear components of $(\partial T/\partial \epsilon)_S$ are illustrated in Fig. \ref{fig-schematic}(d). This suggests that by examining $(\partial T/\partial \epsilon)_S$ over an extended $\epsilon$-range, the constant/linear (even/odd) responses of $dT/d\epsilon$ can be used to distinguish contributions from symmetric and antisymmetric strains. We note that as the out-of-plane mode $\epsilon_{zz}$ transforms identically with $\frac{1}{2}(\epsilon_{xx}+\epsilon_{yy})$ under $C_4$, the above arguments about $\frac{1}{2}(\epsilon_{xx}+\epsilon_{yy})$ also applies to $\epsilon_{zz}$. Before applying this constant/linear criteria to our experimental observations, we note that we have adopted a number of assumptions here, including (1) that the different symmetry channels are independent (\textit{i.e.} higher order terms like $\epsilon_{A_{1g}}\epsilon^2_{B_{1g}}$ are not considered here), (2) that  $C_\epsilon$ doesn't depend strongly on $\epsilon$, which we return to below.

\section{\NoCaseChange{Elastocaloric effect in \ce{DyB2C2}:\\symmetric and antisymmetric strain-tuning of the antiferroquadrupolar (AFQ) phase transition}}

\begin{figure}[t]
	\includegraphics[width =  \columnwidth]{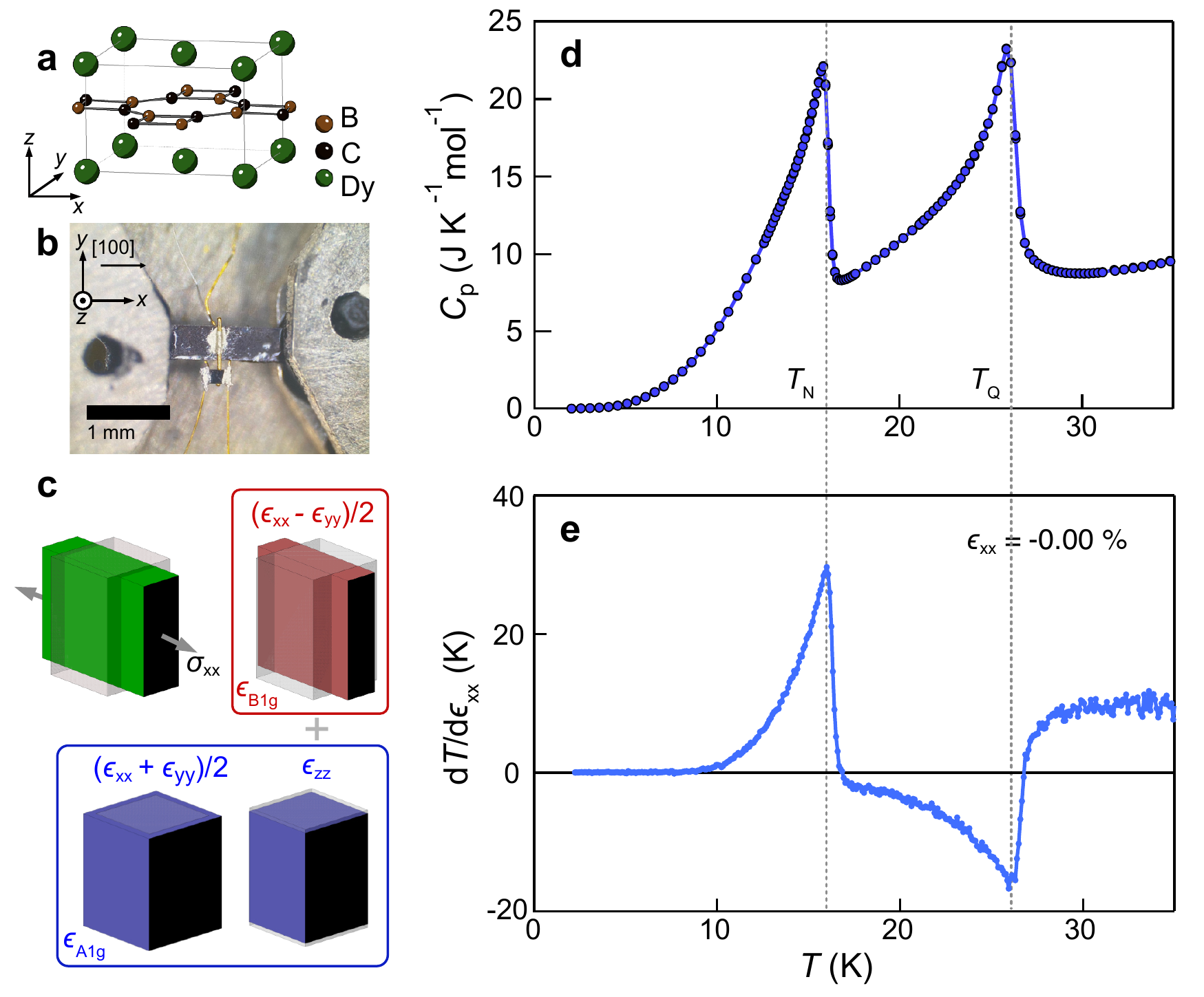}
	\caption{\label{fig-DyB2C2-intro} \textbf{Elastocaloric effect, quadrupolar and magnetic orders in \ce{DyB2C2}} (a) Crystal structure of \ce{DyB2C2} where Dy atoms are shown in green, B atoms in brown and C atoms in black. The bonds between B and C atoms are also depicted. (b) Optical image of a piece of \ce{DyB2C2} single crystal mounted on a strain cell (see text). The scale bar stands for 1 mm. (c) Decomposition of the experimentally generated deformation (green cuboid) into symmetric $\epsilon_{A_{1g}}$ (blue cuboid) and antisymmetric $\epsilon_{B_{1g}}$ (red cuboid) strain channels. The gray box indicate the undeformed tetragonal structure. (d,e) Heat capacity $C_p$ (d) and $dT/d\epsilon_{xx}$ measured without bias strain (e) of \ce{DyB2C2}. The location of $T_Q$ and $T_N$ are indicated by gray dashed lines in (d) and (e).}
\end{figure}

Having outlined the symmetry properties of the elastocaloric effect, we apply the above framework to investigate the strain-response of a tetragonal rare earth intermetallic \ce{DyB2C2}. \ce{DyB2C2} crystallizes in the space group $P4/mbm$; as shown in Fig. \ref{fig-DyB2C2-intro}(a) the Dy square lattice layers are spaced by planar networks of B-C octagon and parallelogram motifs. In Fig. \ref{fig-DyB2C2-intro}(b) we show an optical image of the experimental setup, where a piece of \ce{DyB2C2} single crystal shaped into a long, thin plate is mounted on a strain cell and uniaxial stress is applied along the in-plane [100] axis (which we denote as $x$). The temperature oscillation generated by the AC strain via the elastocaloric effect is measured using the thermometer/thermocouple attached to the surface of the sample, and the AC strain frequency is selected such that the system approaches a quasi-adiabatic condition (see Supplementary Materials). In this experiment we control and measure $\epsilon_{xx}$ with the applied $\sigma_{xx}$, while $\epsilon_{zz}$ and $\epsilon_{yy}$ are left unconstrained ($\sigma_{yy}=\sigma_{zz}=0$); based on Poisson ratios $\nu_{12}=-d\epsilon_{yy}/d\epsilon_{xx}=0.45$ and $\nu_{13}=-d\epsilon_{zz}/d\epsilon_{xx}=0.08$ estimated from the elastic moduli tensor of \ce{LuB2C2} \cite{LuB2C2} (see Supplementary Materials), we linearly decompose the induced deformation into its components in symmetric and antisymmetric channels with their relative strength as $\frac{1}{2}(\epsilon_{xx}-\epsilon_{yy}):\frac{1}{2}(\epsilon_{xx}+\epsilon_{yy}):\epsilon_{zz}=0.72:0.28:-0.08$ (these numbers are normalized with respect to $\epsilon_{xx}$). The decomposition is schematically illustrated in Fig. \ref{fig-DyB2C2-intro}(c): the experimentally generated deformation is shown as a green cuboid and its symmetric $A_{1g}$ (antisymmetric $B_{1g}$) component as blue (red) cuboids. We report the measured elastocaloric coefficient as the oscillation amplitude of $T$ normalized by that of $\epsilon_{xx}$ which we experimentally control (therefore the measured quantity can be expressed as $\dfrac{dT}{d\epsilon_{xx}}\Bigr|_{\bm{\sigma}=\sigma_{xx}}$); hereafter we denote this as $dT/d\epsilon_{xx}$ unless otherwise stated. We note that $dT/d\epsilon_{xx}$ thus defined contains both $\epsilon_{A_{1g}}$- and $\epsilon_{B_{1g}}$-components.

The Dy sites in \ce{DyB2C2} possess a tetragonal local symmetry and the crystal field ground state of \ce{Dy^{3+}} ($4f^9$) is a quasi-quartet $J_z\approx|\pm\frac{1}{2}\rangle,|\pm\frac{3}{2}\rangle$ within the $J=\frac{15}{2}$ manifold \cite{DyB2C2_orbital}. Here $J$ is the total angular momentum operator. This $f$-electron quartet provides a basis for a two-step phase transition to fully release the $R\ln4$ entropy ($R$ is the gas constant) and an avenue to order in a time-reversal-symmetric quadrupolar channel (order parameter characterized by $O_{x^2-y^2}\equiv J_x^2-J_y^2$ and $O_{xy}\equiv J_xJ_y+J_yJ_x$) prior to magnetic order  (order parameter characterized by $J_x,J_y,J_z$). In the quadrupolar order that sets in at about 25 K, the charge distribution of the $f$ electrons breaks the high temperature local four-fold rotational symmetry and the resulting quadrupole moments (linear superposition of $O_{x^2-y^2}$ and $O_{xy}$) in neighboring layers are orthorgonal to each other, forming an antiferroquadrupole (AFQ) state \cite{DyB2C2_yamauchi1999,DyB2C2_Xray}. In this AFQ state, a combined $C_4$ and translational symmetry (translation operation performed along $c$) is preserved, analogous to the combined time-reversal and translational symmetry of a N\'{e}el order. At 16 K an additional phase transition into a non-collinear canted antiferromagnetic (CAFM) phase takes place; there magnetic moments develop within the $ab$-plane and approximately perpendicular to the charge clouds, resulting in a net magnetization along the \{100\} axes \cite{zaharko2004quadrupolar,DyB2C2_yamauchi1999}. The two phase transitions can be traced in the heat capacity $C_p$ (measured using a relaxation method) in Fig. \ref{fig-DyB2C2-intro}(d). We define the higher and lower $T$ anomalies as $T_Q$ (AFQ) and $T_N$ (CAFM), respectively; both phase transitions have been reported to be of second-order nature \cite{DyB2C2_yamauchi1999}.

In Fig. \ref{fig-DyB2C2-intro}(e) we show $dT/d\epsilon_{xx}$ taken at nominally zero bias strain $\epsilon_{xx}=0$. We observe two anomalies with opposite signs in the $T$-dependence of $dT/d\epsilon_{xx}$, corresponding to $T_Q$ and $T_N$ in $C_p$ in Fig. \ref{fig-DyB2C2-intro}(d). The qualitative resemblance of the singularities in $dT/d\epsilon_{xx}$ and $C_p$ is related to the fact that both $C_p$ and $dT/d\epsilon_{xx}$ are second-order derivatives of thermodynamic potentials. In analogy to the Ehrenfest relation in thermal expansion at second order phase transitions \cite{elastic_1975}, we derive a modified Ehrenfest relation for $dT/d\epsilon_{xx}$ (see Appendix A):
\begin{equation}\label{EC_PT}
\dfrac{dT_C}{d\epsilon_{xx}}=\dfrac{\Delta[C(dT/d\epsilon_{xx})]}{\Delta C} 
\end{equation}
where the evolution of critical temperature $T_C$ with strain $dT_C/d\epsilon_{xx}$ is related to the jump of $C(dT/d\epsilon_{xx})$ along with that in the heat capacity $C$ at the phase transition. From $dT/d\epsilon_{xx}$ shown in Fig. \ref{fig-DyB2C2-intro}(e) and assuming $C_p$ in Fig. \ref{fig-DyB2C2-intro}(d) provides a reasonable approximation for $C$ in Eq. \ref{EC_PT} given the small compressibility of solids, we estimate $dT_N/d\epsilon_{xx}=45$ K and $dT_Q/d\epsilon_{xx}=-35$ K near $\epsilon_{xx}=0$. 

\begin{figure}[t]
	\includegraphics[width = \columnwidth]{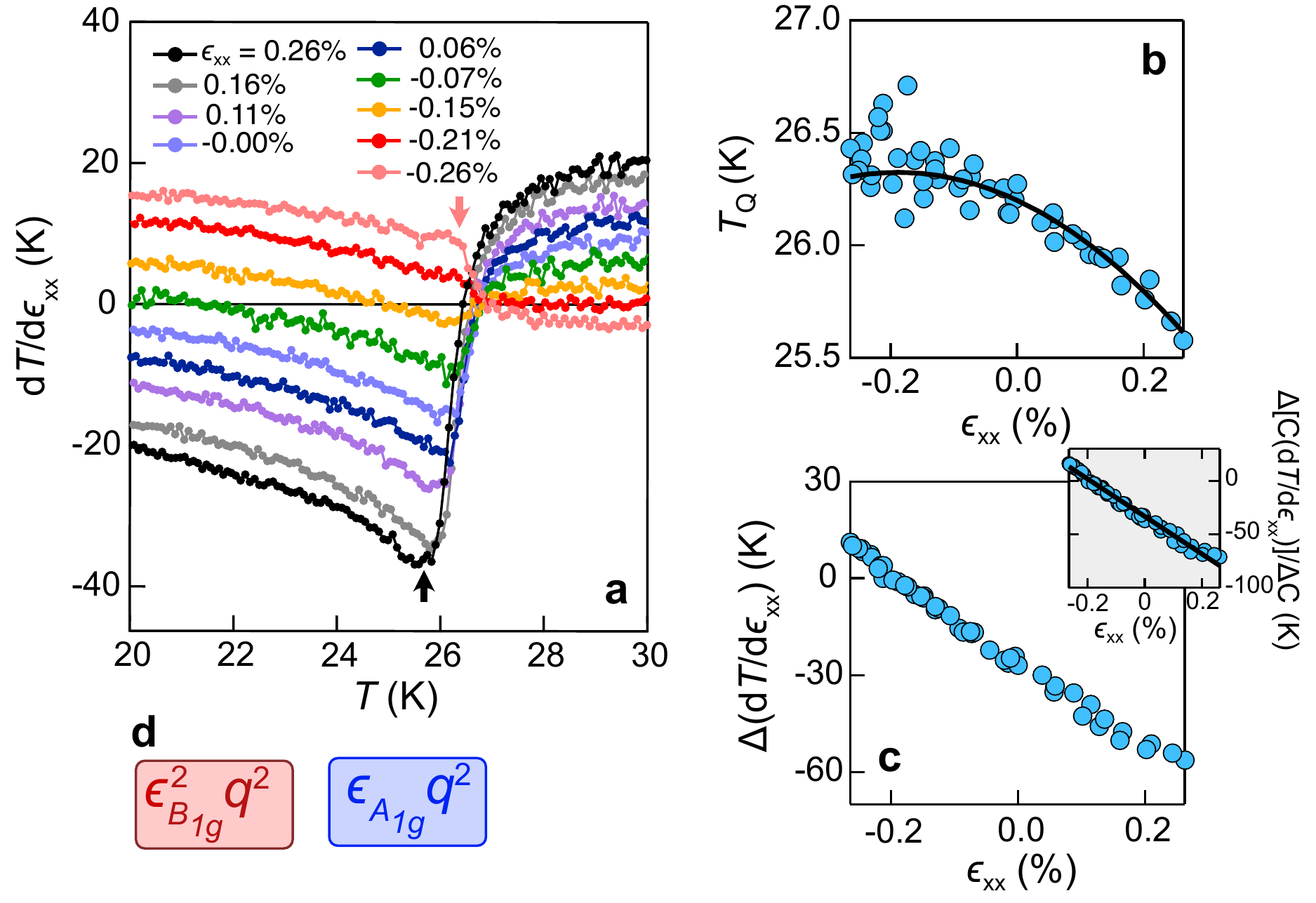}
	\caption{\label{fig-AFQ}\textbf{Strain tuning of the antiferroquadrupolar (AFQ) phase transition in \ce{DyB2C2}} (a) $T$-dependence of $dT/d\epsilon_{xx}$ at selected bias strain $\epsilon_{xx}$.  At $\epsilon_{xx}=\pm0.26\%$, we highlight the discontinuities at $T_Q$ with arrows. (b,c) Evolution of $T_Q$ (b) and $dT/d\epsilon_{xx}$ jump $\Delta (dT/d\epsilon_{xx})\equiv dT/d\epsilon_{xx}(T=T_Q)-dT/d\epsilon_{xx}(30\,\mathrm{K})$ (c) with $\epsilon_{xx}$. Inset of (c) shows $\Delta[C(dT/d\epsilon_{xx})]/\Delta C$ at $T_Q$. The solid curve in (b) is inferred from the linear fit shown as black solid line in (c) inset (see text). (d) Leading coupling between $\epsilon_{A_{1g}}$ ($\epsilon_{B_{1g}}$) and the staggered quadrupolar order parameter $q$ in the free energy enclosed in a blue (red) box. }
\end{figure}

Having demonstrated that $dT/d\epsilon_{xx}$ can be used to trace both AFQ and CAFM phase transitions in \ce{DyB2C2}, we proceed to the evolution of $dT/d\epsilon_{xx}$ under bias $\epsilon_{xx}$. We first focus on the higher $T$ AFQ phase transition and show the $T$-dependence of $dT/d\epsilon_{xx}$ between $20\,$K and $30\,$K at selected $\epsilon_{xx}$ in Fig. \ref{fig-AFQ}(a). A non-zero $\epsilon_{xx}$ introduces a number of changes which we outline as follows. Firstly, the location of $T_Q$ (marked by arrows in Fig. \ref{fig-AFQ}(a)) moves with $\epsilon_{xx}$, and the trend is summarized in Fig. \ref{fig-AFQ}(b). Secondly, at $T_Q$, a deep, negative jump at the most tensile $\epsilon_{xx}$ gradually evolves into a positive, step-like feature at the most compressive $\epsilon_{xx}$. We note that there are intermediate $\epsilon_{xx}$ values (between -0.15\% and -0.21\%) where the singularity associated with $T_Q$ appears vanishingly small; the $\epsilon_{xx}$-evolution of the jump at $T_Q$ of $dT/d\epsilon_{xx}$  (\textit{i.e.} $\Delta(dT/d\epsilon_{xx})$, estimated as $dT/d\epsilon_{xx}(T=T_Q)-dT/d\epsilon_{xx}(30\,\mathrm{K})$) is summarized in Fig. \ref{fig-AFQ}(c). Finally, away from the phase transitions both above and below $T_Q$, we observe a continuous variation of $dT/d\epsilon_{xx}$ with $\epsilon_{xx}$; we return to these behaviors in section V. 

Eq. \ref{EC_PT} suggests that given $dT/d\epsilon_{xx}$ traces at various bias $\epsilon_{xx}$, two independent means exist to trace the strain-dependence of a phase transition: either focusing on the singularity in $dT/d\epsilon_{xx}$ marking $T_Q$, or the size of jump in $dT/d\epsilon_{xx}$ in conjunction with $C$ at $T_Q$. The former and latter views of the AFQ phase transition are contrasted in Fig. \ref{fig-AFQ}(b) and inset of (c), respectively. In the latter we assume that $C$ for both the ordered and disordered phases near $T_Q$ are independent of $\epsilon$. In Fig. \ref{fig-AFQ}(b), $T_Q$ appears to evolve with $\epsilon_{xx}$ both in a linear $\epsilon_{xx}$ and quadratic $\epsilon_{xx}^2$ manner. Alternatively, $\Delta[C(dT/d\epsilon_{xx})]/\Delta C$ shown in Fig. \ref{fig-AFQ}(c) inset appears linear with $\epsilon_{xx}$ over the entire $\epsilon_{xx}$ range, and a linear fit $A+B\epsilon_{xx}$ yields $A=-47.1(7), B=-2.49(4)\times10^4$. Following Eq. \ref{EC_PT}, $\Delta[C(dT/d\epsilon_{xx})]/\Delta C$ should reflect the $\epsilon_{xx}$-evolution of ${dT_Q}/{d\epsilon_{xx}}$: in Fig. \ref{fig-AFQ}(b) we compare an integration of $A+B\epsilon_{xx}$ multiplied by a factor to account for the imperfect adiabaticity in our experiments (black solid curve) with the directly traced $T_Q(\epsilon_{xx})$ (blue symbols) \footnote{The black solid curve is scaled along the temperature axis by a factor of 4, which we hypothesize originates from an imperfect adiabaticity of our experiments.}. That the integrated curve agrees with the bare $\epsilon_{xx}$-dependence of $T_Q$ provides an experimental demonstration of the validity of the modified Ehrenfest relation Eq. \ref{EC_PT} and additionally suggests that our assumption that the heat capacity is not strongly affected by $\epsilon$ is valid. We note that a direct polynomial fit to $T_Q(\epsilon_{xx})$ yields uncertainties of the $\epsilon_{xx}$($\epsilon_{xx}^2$)-coefficients to be 6\%(27\%); thanks to its strain-derivative nature, the elastocaloric coefficient at $T_Q$ provides with significantly improved accuracy (1-2\% for both $\epsilon_{xx}$ and $\epsilon^2_{xx}$ terms) a view of the evolution of $T_Q$ with strain. 

Following the symmetry-decomposition framework we laid out in section II, we assign the $\epsilon_{xx}$ and $\epsilon_{xx}^2$ terms in $T_Q$ to the symmetric ($\epsilon_{A_{1g}}$) and antisymmetric ($\epsilon_{B_{1g}}$) strain channels, respectively. They are expected to result from leading terms in the free energy of the form $\epsilon_{A_{1g}}q^2$ ($\epsilon_{B_{1g}}^2q^2$) for $\epsilon_{A_{1g}}$ ($\epsilon_{B_{1g}}$), as shown in Fig. \ref{fig-AFQ}(d) ($q$ represents the staggered quadrupolar order parameter). The absence of an $\epsilon_{A_{1g}}^2$ term in $T_Q$ is corroborated by the linear pressure-evolution of $T_Q$ up to hydrostatic pressure of 8 GPa, where the in-plane $\epsilon_{A_{1g}}$ component $\frac{1}{2}(\epsilon_{xx}+\epsilon_{yy})$ reaches $-0.8\%$ \cite{ikeda2018symmetric,DyB2C2_pressure}. The above symmetry decomposition allows us to conclude that antisymmetric strain suppresses $T_Q$ in a quadratic manner, and that at the highest strain in our experiments (both compressive and tensile), the antisymmetric strain contribution ($\epsilon_{xx}^2$ term) becomes comparable and exceeds the symmetric strain contribution ($\epsilon_{xx}$ term).

\section{\NoCaseChange{Tetragonal-symmetry-breaking in the canted antiferromagnetic (CAFM) phase}}

\begin{figure}[t]
	\includegraphics[width = \columnwidth]{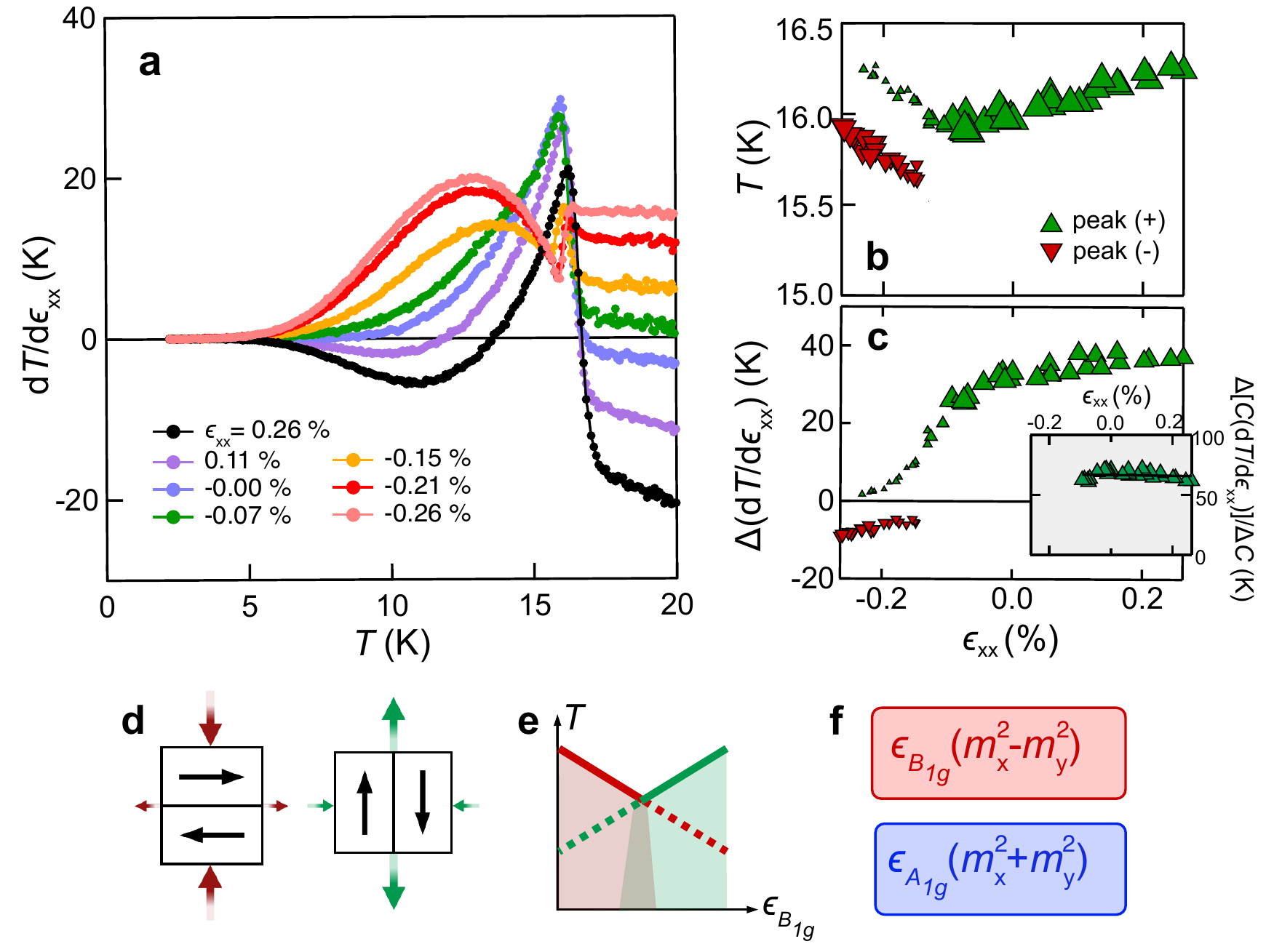}
	\caption{\label{fig-AFM}\textbf{Strain tuning of the canted antiferromagnetic (CAFM) phase transition in \ce{DyB2C2}} (a) $dT/d\epsilon_{xx}$ at selected $\epsilon_{xx}$ below 20 K. (b) $T-\epsilon_{xx}$ phase diagram for the CAFM phase. Green and red triangle symbols are extracted from locations of positive and negative peaks in $dT/d\epsilon_{xx}(T)$, respectively. (c) $\Delta(dT/d\epsilon_{xx})$ as a function of $\epsilon_{xx}$. The size of the symbols in (b) and (c) is proportional to natural logarithm of the peak width determined from zeros of the $T$-derivative of $dT/d\epsilon_{xx}$. Inset of (c) shows $\Delta[C(dT/d\epsilon_{xx})]/\Delta C$ of the positive peaks and the solid black line depicts a linear fit. (d) Schematic of the two sets of the four-fold CAFM domains; black arrows indicate the orientation of net magnetization $m$, and thick colored arrows illustrate the strain that selects these domains. (e) Tuning of $T_N$ of the two domains with $\epsilon_{B_{1g}}$. The solid (dashed) lines indicate $T_N$ of the favored (unfavored) domain. (f) Coupling terms between $\epsilon_{B_{1g}}$ and $\epsilon_{A_{1g}}$ to the CAFM net moment $m$ to the leading order: $\epsilon_{B_{1g}}(m_x^2-m_y^2)$ (red box) and  $\epsilon_{A_{1g}}(m_x^2+m_y^2)$ (blue box). }
\end{figure}

Having identified considerable contributions from both $\epsilon_{A_{1g}}$ and $\epsilon_{B_{1g}}$ in the strain-modulation of $T_Q$, we turn to the lower CAFM phase transition. In Fig. \ref{fig-AFM}(a), starting from $\epsilon_{xx}$ on the tensile side $\epsilon_{xx}=0.26\%$ (black curve), the peak in $dT/d\epsilon_{xx}$ at $T_N$ sits on top of a continuously varying background, and its height is initially not much varied by decreasing $\epsilon_{xx}$; below $\epsilon_{xx}=-0.1\%$, however, a downward peak resembling a horizontal mirror image of the peak at $\epsilon_{xx}>0$ abruptly emerges. This can be contrasted with the continuous variation of the phase transition jumps at $T_Q$ in Fig. \ref{fig-AFQ}(a). In Fig. \ref{fig-AFM}(b) and (c) we trace the location of these peaks in $dT/d\epsilon_{xx}$ and the height of the associated jumps $\Delta(dT/d\epsilon_{xx})$, respectively. Due to the presence of multiple singular features at $\epsilon_{xx}=-0.2\sim-0.13\%$, in Fig. \ref{fig-AFM}(b,c) we mark the peaks with the respective symbol sizes proportional to their peak widths (see Supplementary Materials). 

In Fig. \ref{fig-AFM}(b) we can identify a ``V-shape'' in the $T-\epsilon_{xx}$ plane with its two branches marked by the primary positive and negative peak features, respectively; this is accompanied by a step-wise structure of $\Delta(dT/d\epsilon_{xx})$ as a function of $\epsilon_{xx}$ displayed in Fig. \ref{fig-AFM}(c). The ``branched" phase boundary structure prompts us to propose the emergence of two distinct sets of domains below $T_N$ that are strain-selective. Above $\epsilon_{xx}=-0.13\%$, a linear fit to $\Delta[C(dT/d\epsilon_{xx})]/\Delta C$ (Fig. \ref{fig-AFM}(c) inset) yields $A=66.6(8),B=-3(6)\times10^2$ ($A+B\epsilon_{xx}$), which via Eq. \ref{EC_PT} suggests a linear dependence of $T_N$ with $\epsilon_{xx}$, consistent with our observation (Fig. \ref{fig-AFM}(b) \footnote{To quantitatively compare $A\epsilon_{xx}$ with $T_N(\epsilon_{xx})$ in Fig. \ref{fig-AFM}(b), an additional factor 2.3 is required in front of $A$. We note that this factor introduced to account for imperfect adiabaticity is comparable with that used above for $T_Q$. The difference between the two factors may arise from a $T$-dependence of the thermal conditions (thus the adiabaticity) of the experimental setup.}; we note that the $B$ coefficient, which via Eq. \ref{EC_PT} is associated with a potential $\epsilon_{xx}^2$ term in $T_N(\epsilon_{xx})$, cannot be distinguished from zero and is at least two orders of magnitude smaller than its counterpart in $T_Q(\epsilon_{xx})$.

Below we discuss the nature of the strain-selective domains and the linear evolution of $T_N$ with $\epsilon$. In the CAFM phase of \ce{DyB2C2}, it is known that there exists net magnetization $m$ along in-plane \{100\} directions \cite{DyB2C2_yamauchi1999,zaharko2004quadrupolar}; via magnetoelastic coupling \cite{magnetism2002}, these domains can distort along \{100\} below $T_N$, providing a natural origin for the proposed strain-selective domains. The lowest order magnetoelastic coupling is required to take the form of $\epsilon m^2$ to be invariant under time-reversal symmetry \cite{magnetism2002}, and we group the four \{100\} magnetic domains into subgroups with $m$ along $\pm x$ and $\pm y$, as illustrated in Fig. \ref{fig-AFM}(d). Additionally taking into account $C_4$ (for discussions hereafter, we still adopt the convention of the high $T$ $D_{4h}$ group), we may decompose $\epsilon m^2$ into terms containing antisymmetric $\epsilon_{B_{1g}}(m_x^2-m_y^2)$ and symmetric strain $\epsilon_{A_{1g}}(m_x^2+m_y^2)$, respectively \cite{fernandes2012manifestations}. We note that the consequences of the $\epsilon_{B_{1g}}$ term are two-fold: it selects among the differently distorted domains, and promotes the order within the favored domains while suppressing that of the other type (both in an $\epsilon$-linear fashion), as we illustrate in Fig. \ref{fig-AFM}(e); joining $T_N(\epsilon_{B_{1g}})$ of the favored domains yields a V-shape, akin to that observed in Fig. \ref{fig-AFM}(b). We note that similar phase diagrams have been discussed for antisymmetric strain manipulation of in-plane charge density wave orders in tetragonal \ce{(Er,Tm)Te3} \cite{RTe3} and in the context of $p_x\pm ip_y$-type multi-component superconducting phases \cite{hicks2014strong}, both with similar forms of order parameter-antisymmetric strain coupling. 

\begin{figure}[t]
	\includegraphics[width = \columnwidth]{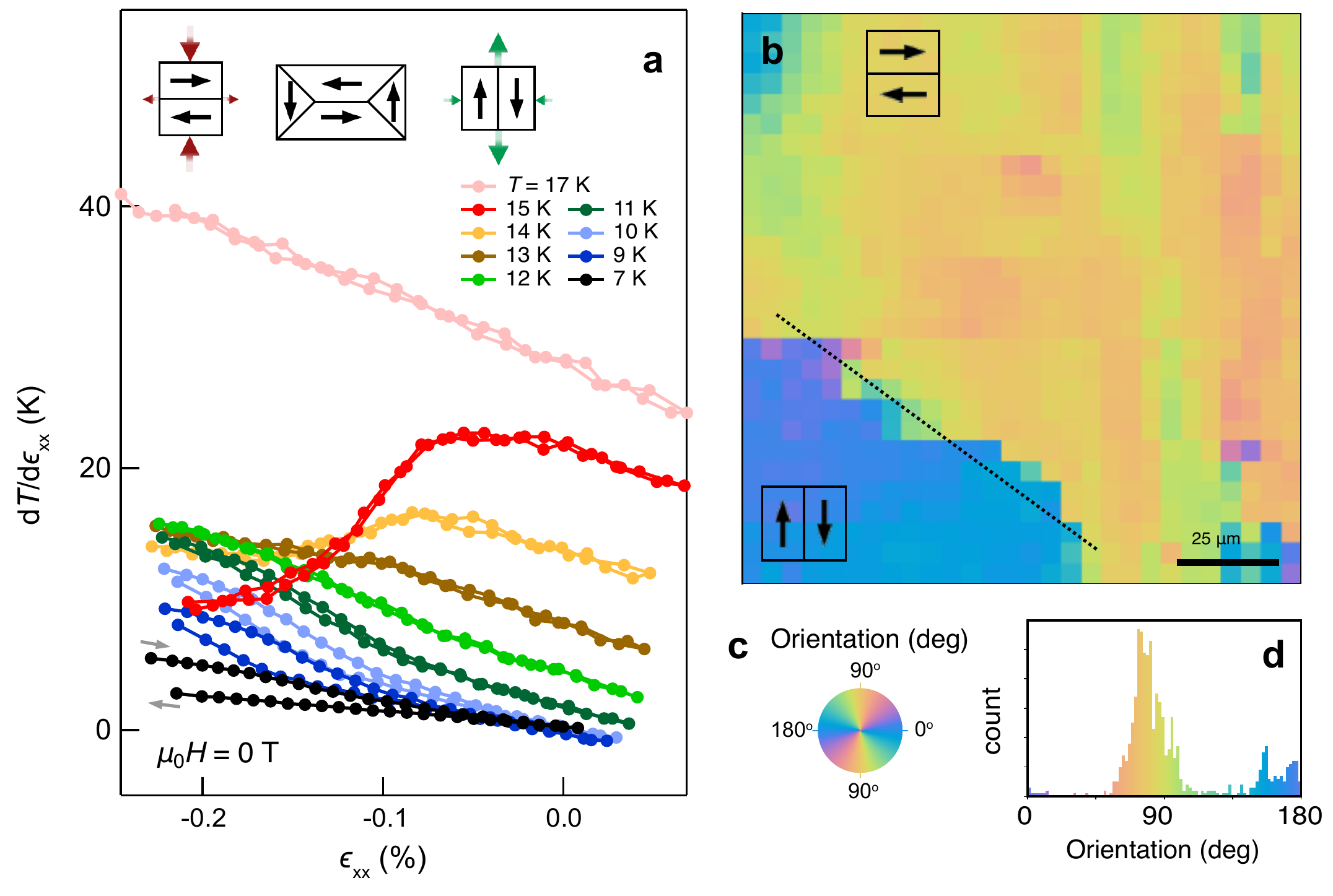}
	\caption{\label{fig-AFM-orthor}\textbf{Tetragonal symmetry-breaking in the CAFM phase of \ce{DyB2C2}} (a) $\epsilon_{xx}$-hysteresis of $dT/d\epsilon_{xx}$ at selected $T$ at zero magnetic field. The schematics at the top illustrate the favored domains on the compressive (left) and tensile (right) strain side, respectively. The pink curve ($T=17$ K) is offset by 30 K for clarity. The gray arrows indicate the direction of the $\epsilon_{xx}$-scans in the hysteresis. (b) Map of principal axis orientation over a $130 \times 130 \mathrm{\mu m}^2$ area with step size of $5\mathrm{\mu m}$ with the orientation defined in (c). Dashed line is a guide to the eye outlining an observed domain boundary. (d) Histogram of orientations shown in (b), indicating two domains whose optic axes are offset by $90^{\circ}$ from one another. Measurements in (b-d) were taken at 2.5 K. }
\end{figure}

In Fig. \ref{fig-AFM-orthor}(a) we show $dT/d\epsilon_{xx}$ taken in strain scans at selected $T$: above $T_N$ $dT/d\epsilon_{xx}$ appears linear in $\epsilon$ over the measured strain range, while below $T_N$ a non-linear component develops; below $12\,$K a strain-hysteresis additionally opens up. Both non-linearity and hysteresis in $dT/d\epsilon_{xx}(\epsilon_{xx})$ are consistent with $\epsilon$-selectivity of the CAFM domains: we attribute the positive strain limit where the system reaches a constant slope to a state with a single type of orientational domains, which is separated by a multi-domain state from a state with the alternative set of orientational domains (see insets of Fig. \ref{fig-AFM-orthor}(a)). The domain-selection necessarily arises from antisymmetric strain and requires an $\epsilon_{B_{1g}}$ lattice distortion in each of the domains. 

The symmetry-lowering at the CAFM phase is additionally evidenced by the mapping of thermally modulated optical birefringence shown in Fig. \ref{fig-AFM}(b-d); a similar technique had been used to image nematic domains in other systems \cite{little2020three} (Additional information on the optical setup can be found in Supplementary Materials). In Fig. \ref{fig-AFM}(b) two types of domains with their principal axes near the crystallographic \{100\} orientations (Fig. \ref{fig-AFM}(c,d)) can be identified, and their domain wall appears to run approximately 45$^{\circ}$ between the principal axes. The birefringence imaging confirms the broken $C_4$ with an orthorhombicity developing along \{100\} below $T_N$, and complements the above thermodynamic evidences of tetragonal symmetry-breaking in the CAFM phase. The observed diagonal domain wall is also consistent with that commonly observed on surfaces of ferromagnets with four-fold easy axes in the plane \cite{hubert2008magnetic}. It has been suggested previously that magnetic order in \ce{DyB2C2} is accompanied by slight rotation of quadrupole moments from a staggered arrangement due to a competition between quadrupolar exchange and magnetic exchange interactions, and as a result the crystalline symmetry is lowered from tetragonal to orthorhombic \cite{zaharko2004quadrupolar}; that the CAFM phase is potentially monoclinic is proposed in a thermal expansion study \cite{DyB2C2_thermalexpansion}. Our experimental observations at present cannot rule out a monoclinic nature of the CAFM phase, while we can set a bound that the highest point group symmetry of the CAFM phase is orthorhombic with the in-plane principal axes along the high temperature tetragonal \{100\} orientations.

\section{\NoCaseChange{Anti-symmetric elastocaloric effect from staggered quadrupole moments}}
\begin{figure}[t]
	\includegraphics[width =  \columnwidth]{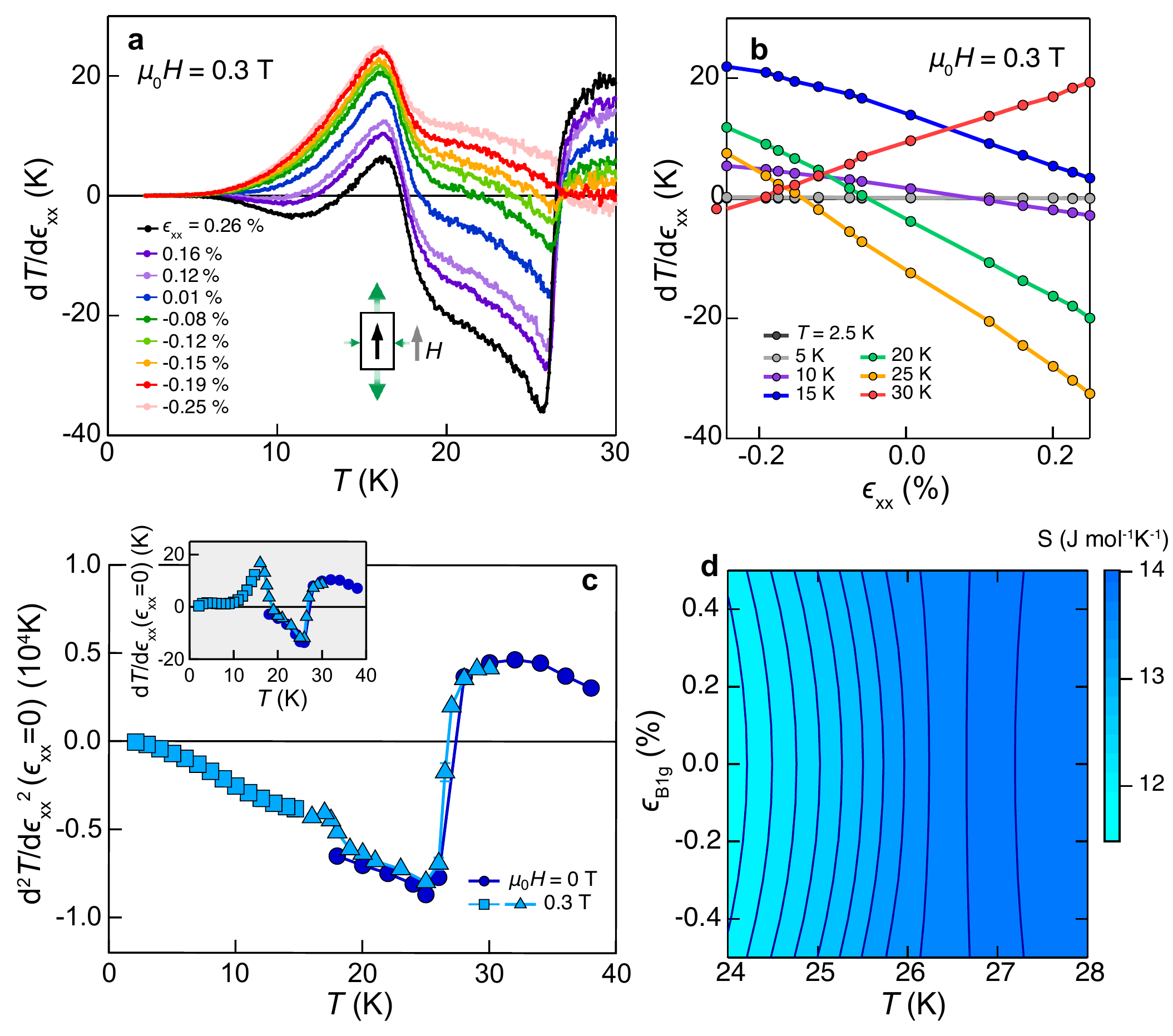}
	\caption{\label{fig-linear}\textbf{Antisymmetric elastocaloric effect in \ce{DyB2C2}} (a) $dT/d\epsilon_{xx}$ at $\epsilon_{xx}$ in an applied magnetic field $H$ along [100] of 0.3 T. Inset shows a schematic of the orientation of $H$, the selected magnetic domain, and an applied tensile strain ($\epsilon_{xx}>0$). (b) Constant $T$-cuts of $dT/d\epsilon_{xx}$ at selected $T$. (c) $\epsilon_{xx}$-slope of $dT/d\epsilon_{xx}$ as a function of $T$. Inset shows the intercept at $\epsilon_{xx}=0$. Light (dark) blue symbols are linear fit parameters extracted at 0.3 T (0 T). The triangular and circular symbols are extracted from data take at a fixed strain frequency of 27.1 Hz while square symbols from data taken at adjusted strain frequencies with $T$ (see Supplementary Materials).  (d) Entropy landscape with respect to $\epsilon_{B_{1g}}$ inferred from (c) (see Supplementary Materials) near $T_Q$.}
\end{figure}

In the preceding sections we focused on the strain tuning of $T_Q$ and $T_N$; in the following we examine the overall strain responses in the system not limited to the proximity of the phase transitions. The intertwined magnetic and quadrupolar degrees of freedom motivates us to use a small polarizing magnetic field to eliminate complications from hysteretic and multi-domain behaviors. $dT/d\epsilon_{xx}$ obtained with a magnetic field of 0.3 T parallel to [100] (also the stress direction) are summarized in Fig. \ref{fig-linear}(a) (see Supplementary Materials for $M(H)$ hysteresis of an unstrained sample). We note that the application of $H$ appears to simplify $dT/d\epsilon_{xx}$ near $T_N$ as compared to the zero field responses (\textit{i.e.} the peak does not reverse in sign as a function of $\epsilon_{xx}$, in contrast to the zero field cases seen in Fig. \ref{fig-AFM}(a) and Fig. S6) and that the peak at $T_N$ in $H$ is rounded due to the presence of a finite net magnetization. $dT/d\epsilon_{xx}$ at different $\epsilon_{xx}$ in Fig. \ref{fig-linear}(a) appear to be composed of an $\epsilon_{xx}$-independent contribution to $dT/d\epsilon_{xx}$ (which can be approximated by the blue trace at $\epsilon_{xx}=0.01\%$) and a component that varies monotonically with $\epsilon_{xx}$. In Fig. \ref{fig-linear}(b) we examine constant-$T$ $\epsilon_{xx}$ cuts of $dT/d\epsilon_{xx}$. The $\epsilon_{xx}$-linear dependence of $dT/d\epsilon_{xx}$ over the entire strain-range at all $T$ in Fig. \ref{fig-linear}(b) can be contrasted with the nonlinear $dT/d\epsilon_{xx}$ in Fig. \ref{fig-AFM-orthor}(a), and is consistent with a single domain state below $T_N$ in field. 

Results of linear fits to $\epsilon_{xx}$-cuts of $dT/d\epsilon_{xx}$ at $0.3\,$T are summarized in Fig. \ref{fig-linear}(c); in Fig. \ref{fig-linear}(c) we also include both the intercept and slope extracted from zero field $dT/d\epsilon_{xx}$ above $T_N$ as dark blue circles: the close comparison between $0\,$T and $0.3\,$T responses above $T_N$ is consistent with the time-reversal-symmetric nature of the quadrupolar order. That the $T$-trace of the intercept of $dT/d\epsilon_{xx}$ (Fig. \ref{fig-linear}(c) inset) compares closely to the responses at $T_N$ on the tensile strain side at zero field (Fig. \ref{fig-AFM}(a)) suggests that the magnetic domain favored by $H$ coincides with that favored by tensile $\epsilon_{xx}$, from which we infer that the long axis of the distorted unit cell is along $m$ (Fig. \ref{fig-linear}(a) inset) \footnote{We hypothesize that the pre-selection of one set of the domains at nominal zero bias strain may result from a differential thermal contraction between the strain cell and the \ce{DyB2C2} sample.}. Hereafter we focus on the $\epsilon_{xx}$-slope of $dT/d\epsilon_{xx}$ (main panel of Fig. \ref{fig-linear}(c)). As we invoke above, the $\epsilon$-odd component in $dT/d\epsilon$ by symmetry originates from anti-symmetric strain (Fig. \ref{fig-schematic}(g)), which in the present case is $\epsilon_{B_{1g}}$: viewed alongside with Eq. \ref{dSde}, above $T_Q$, $d^2T/d\epsilon^2>0$ indicates a quadratic decrease of $S$ with $\epsilon_{B_{1g}}$, consistent with strain suppression of para-quadrupolar fluctuations; similar $\epsilon$-dependence of $S$ has been reported in iron-based superconductors above the nematic phase transition and attributed to a bilinear coupling between antisymmetric strain and underlying nematic fluctuations \cite{ikeda_elastocaloric_2021}. Below $T_Q$, $d^2T/d\epsilon^2<0$ implies on the contrary a quadratic increase of $S$ with $\epsilon_{B_{1g}}$. In Fig. \ref{fig-linear}(d), we show a contour plot of the entropy landscape with $\epsilon_{B_{1g}}$ inferred from the slope in Fig. \ref{fig-linear}(c) and the zero strain heat capacity (see Supplementary Materials for the procedure to extract $S(\epsilon,T)$ and thus deduce $C(\epsilon,T)$ \cite{Sr2RuO4_ECE}) near $T_Q$, where a curvature change in the entropy landscape across $T_Q$ is apparent. 

\begin{figure}[t]
	\includegraphics[width =  \columnwidth]{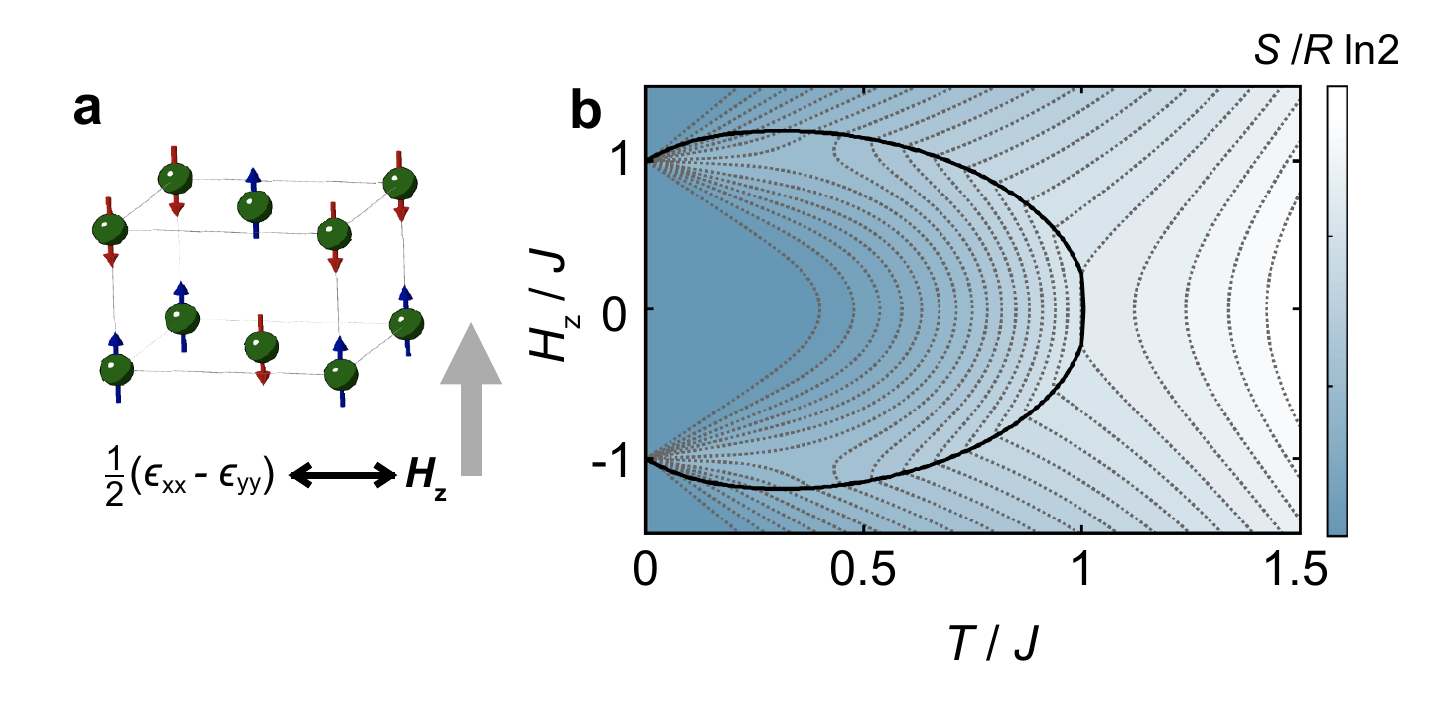}
	\caption{\label{fig-theory}\textbf{Antiferromagnetic (AFM) Ising model} (a) Schematic of an effective AFM Ising model where an antiferromagnetic spin arrangement correspond to the antiferroquadrupole order in \ce{DyB2C2} and the effective magnetic field $H_z$ corresponds to $\epsilon_{B_{1g}}$ (see text). (b) Entropy landscape of the AFM Ising model in the $H_z-T$ plane. The black solid curve in (b) marks the phase boundary between the AFM and paramagnetic phases. An $H_z$-independent phonon contribution is also included in order to obtain (b).}
\end{figure}

To shed light on the antisymmetric strain effects on the AFQ order and the underlying staggered quadrupolar arrangement, we introduce the following Hamiltonian 
\begin{equation}\label{HamiltonianQ}
H=K\sum_{<i,j>}O_iO_j-g\epsilon_{B_{1g}}\sum_iO_i 
\end{equation}
where the first term describes the quadrupole-quadrupole interaction between the nearest neighbors ($K>0$ gives an AFQ order) and the second term the quadrupole-strain coupling with $g>0$ \cite{gehring1975co,luthi2007physical}. Eq. \ref{HamiltonianQ} can be further mapped onto an effective AFM Ising model with the quadrupole moments $O$ mapping onto Ising spins $S^z$ and $\epsilon_{B_{1g}}$ mapping onto an effective magnetic field $H_z$ $(J>0)$ (Fig. \ref{fig-theory}(a)):
\begin{equation}\label{HamiltonianM}
H=J\sum_{<i,j>}S_i^zS_j^z-H_z\sum_iS_i^z 
\end{equation}
whose mean-field entropy landscape is shown in Fig. \ref{fig-theory}(b) (here we also include a field-independent phonon background, see Supplementary Materials). In Fig. \ref{fig-theory}(b), $H_z$ appears to suppress the critical temperature of the AFM order (similar to $\epsilon_{B_{1g}}$ for the AFQ order), near which the curvature of the isentropic contours exhibits a sharp change, giving rise to a sign reversal of the antisymmetric slope (see Supplementary Materials) similar to that observed experimentally in Fig. \ref{fig-linear}(c,d). The comparison suggests that within the ordered state the antisymmetric strain ($\epsilon_{B_{1g}}$ in this case) destabilizes and therefore suppresses the staggered AFQ order akin to how magnetic field destabilizes a staggered Ising antiferromagnetic order by exciting pseudospin flips (quadrupole flops). The non-monotonic shape of the phase boundary in the $(T,H_z)$ plane near $T=0$ has been discussed before for AFM Ising models and attributed to an order-by-disorder effect \cite{AF-ising-ziman1951antiferromagnetism,AF-ising-fcc}. We note that the effective AFM Ising model has only taken into account the quadrupolar degrees of freedom; the continuity of the observed antisymmetric response above and below $T_N$ aside from a small kink at $T_N$ in Fig. \ref{fig-linear}(c) implies that the overall staggered quadrupole configuration is likely not fundamentally modified (a weak relative re-orientation of the quadrupole moments has been suggested by Ref. \cite{zaharko2004quadrupolar}) by the magnetic order and is the source of the observed $\epsilon$-linear responses in $dT/d\epsilon$.

\section{\NoCaseChange{Discussion and summary}}

In summary, we have employed the AC elastocaloric effect to investigate the strain responses of the $f$-electron antiferroquadrupolar order in \ce{DyB2C2}. The strain-dependence of both the quadrupolar and magnetic phase transitions in the system can be precisely charaterized by the jump of the elastocaloric signals--the former contains both linear and quadratic strain-dependences while the latter remain linear over the explored strain range. While symmetric strain $\epsilon_{A_{1g}}$ always appears to tune both $T_N$ and $T_Q$ in a linear manner, the antisymmetric strain $\epsilon_{B_{1g}}$, which is inaccessible in hydrostatic pressure experiments, plays an indispensable and more versatile role in controlling $T_Q$ and $T_N$: for the AFQ phase, $\epsilon_{B_{1g}}$ suppresses $T_Q$ in a quadratic manner, while for the CAFM order, the primary role of $\epsilon_{B_{1g}}$ is found to be two-fold: domain selection as well as linearly tuning $T_N$. The distinct behavior of the two phase transitions with $\epsilon_{B_{1g}}$ lies in the different forms of coupling between $\epsilon_{B_{1g}}$ and the underlying order parameters $\epsilon_{B_{1g}}^2q^2$ and $\epsilon_{B_{1g}}(m_x^2-m_y^2)$ (Fig. \ref{fig-AFQ}(d) and Fig. \ref{fig-AFM}(f)); in this context, we may in turn use the evolution of critical temperatures with antisymmetric strain to place strong constraints on the spatial/lattice symmetry of the underlying order parameters given a generic phase transition.

From a symmetry perspective, $f$-electron-based quadrupolar orders can be viewed as a close analogue of the nematic phases observed in a number of transition-element-based strongly correlated electron systems \cite{YbRu2Ge2,nematicfluid}. The relatively well localized nature of the electronic degrees of freedom and the strong magnetoelastic coupling of $f$-electrons marks them as model systems to drive quantum phase transitions with strain. For instance, it has been proposed that antisymmetric strain orthogonal to a globally uniform ferroquadrupolar order couples to the latter as an effective transverse field, therefore promoting quantum fluctuations and ultimately driving a quantum phase transition into the Ising nematic order \cite{maharaj2017TRIM}. In the present case of an antiferroquadrupolar order, we demonstrate that the antisymmetric strain suppresses the AFQ phase transition, likely through introducing quadrupole flops (pseudospin flips) as an effective ``longitudinal field" in the AFM Ising model (we note that we do not exclude transverse field-like effects akin to those proposed in Ref. \cite{maharaj2017TRIM}). Extrapolating from the measured $T_Q(\epsilon_{B_{1g}})$ we expect that antisymmetric strain on the order of 3-4\% may be required to completely suppress the AFQ order and drive a quantum phase transition in the present system. Our study provides a proof-of-principle example of employing antisymmetric strain as a means of driving quantum phase transitions in spatially varying anisotropic electronic orders beyond a uniform rotation-symmetry-breaking nematic order; examples of systems to which we can extend the above study include orbital ordering in transition metal oxides \cite{kugel_Khomski,LaMnO3}, spin and charge stripe order in the low temperature tetragonal phase in \ce{La_{2-x}Ba_xCuO4} \cite{LaBaCuO4_PRL,LaBaCuO4_PRB}, along with a few  ``hidden order states" in a number of $f$ electron systems \cite{RevModPhys_Multipole}.

Viewed alternatively from the perspective of employing elastocaloric effect as a tool to study strain responses, our results establish that the elastocaloric coefficients provide a refined picture of strain-evolution of given phase transitions, via a modified Ehrenfest relation thanks to its thermodynamic and strain-derivative nature. Additionally, our case study here demonstrates that extending the elastocaloric measurement over a range of strain values provides a pathway to systematically extract effects from spatially symmetric and antisymmetric strains; we anticipate that this framework can be applied as a powerful organizing principle for exploring strain responses of extended classes of quantum materials.

\begin{acknowledgements}
We thank R.M. Fernandes and A.P. Mackenzie for fruitful discussions. Experimental work performed at Stanford University was funded by the Gordon and Betty Moore Foundation EPiQS Initiative, grant GBMF9068. L.Y. also acknowledges support by the Marvin Chodorow Postdoctoral Fellowship at the Department of Applied Physics, Stanford University. M.D.B. acknowledges support by the Geballe Laboratory for Advanced Materials Fellowship. Optical measurements were performed at the Lawrence Berkeley Laboratory as part of the Quantum Materials program, Director, Office of Science, Office of Basic Energy Sciences, Materials Sciences and Engineering Division, of the U.S. Department of Energy under Contract No. DE-AC02-05CH11231. V.S. is supported by the Miller Institute for Basic Research in Science, UC Berkeley. J.O. and Y.S. received support from the Gordon and Betty Moore Foundation’s EPiQS Initiative through Grant GBMF4537 to J.O. at UC Berkeley. J.F.R.N. acknowledges support from the Gordon and Betty Moore Foundation’s EPiQS Initiative through Grants GBMF4302 and GBMF8686. 
\end{acknowledgements}

\appendix
\section{\NoCaseChange{$\epsilon-T$ Ehrenfest Relation}}

Here we consider in the $\epsilon-T$ plane two phases 1 and 2 that are separated from each other by a second order phase transition with critical temperature $T_C(\epsilon)$. The state variables of the two phases, such as entropy $S_{1,2}$, should be continuous everywhere along the phase boundary $T_C(\epsilon)$ in the $\epsilon-T$ plane ($S_1=S_2\rvert_{T_C}$). Taking an infinitesimal variation of both $S_1$ and $S_2$ along $T_C(\epsilon)$ results in $dS_1=dS_2\rvert_{T_C}$, which requires
\begin{equation}
    \left(\dfrac{\partial S_1}{\partial T}\right)_{\epsilon}dT+\left(\dfrac{\partial S_1}{\partial \epsilon}\right)_{T}d\epsilon=\left(\dfrac{\partial S_2}{\partial T}\right)_{\epsilon}dT+\left(\dfrac{\partial S_2}{\partial \epsilon}\right)_{T}d\epsilon
\end{equation}
along $T_C(\epsilon)$. This suggests that
\begin{equation}
\dfrac{dT_C}{d\epsilon}=\dfrac{(\partial S_1/\partial \epsilon)_{T}-(\partial S_2/\partial \epsilon)_{T}}{(\partial S_2/\partial T)_{\epsilon}-(\partial S_1/\partial T)_{\epsilon}}    
\end{equation}
Using $TdS=C_{\epsilon}dT$ and Eq. (1) we get
\begin{equation}\label{jump_1}
\dfrac{dT_C}{d\epsilon}=\dfrac{C_1(\partial T_1/\partial \epsilon)_{S}-C_2(\partial T_2/\partial \epsilon)_{S}}{C_1-C_2} 
\end{equation}
where $C_1$ and $C_2$ are the heat capacity of the two phases at $T_C$. Eq. \ref{jump_1} may be reformulated as
\begin{equation}\label{Ehrenfest}
\dfrac{dT_C}{d\epsilon}=\dfrac{\Delta[C(\partial T/\partial \epsilon)_{S}]}{\Delta C} 
\end{equation}
analogous to the Ehrenfest relation relating the phase boundary with respect to uniaxial pressure $T_C(P_i)$ to the jumps at thermal expansion $\alpha_i$ and heat capacity \cite{elastic_1975}:
\begin{equation}
\dfrac{dT_C}{dP_i}=V_mT_C\dfrac{\Delta\alpha_i}{\Delta C} 
\end{equation}
Here $V_m$ is the molar volume.

The above derivation is based on a generic form of applied $\epsilon$. Taking into account the experimentally relevant linear combination of strain modes discussed in the main text, the modified Ehrenfest relation in the context of our experiments can be expressed as
\begin{equation}\label{Ehrenfest1}
\dfrac{dT_C}{d\epsilon_{xx}}\Bigr|_{\bm{\sigma}=\sigma_{xx}}=\dfrac{\Delta[C(dT/d\epsilon_{xx})\rvert_{\bm{\sigma}=\sigma_{xx}}]}{\Delta C} 
\end{equation}
where $(dT/d\epsilon_{xx})\rvert_{\bm{\sigma}=\sigma_{xx}}$ is the experimentally measured elastocaloric coefficient.
\bibliography{reference}
\end{document}